\documentclass[10pt,a4paper]{article}
\usepackage[utf8]{inputenc}
\usepackage[english]{babel}
\usepackage{amsmath}
\usepackage{amsfonts}
\usepackage{amssymb}
\usepackage{graphicx}
\usepackage{subfigure}
\usepackage{color}
\usepackage{authblk}
\usepackage{cite}
\usepackage{url}
\usepackage{float}

\title{Phase space structure and escape time dynamics in a Van der Waals model for exothermic reactions}

\author[a]{Francisco Gonzalez Montoya}
\author[a]{Stephen Wiggins}
\affil[a]{ School of Mathematics, University of Bristol, Bristol, BS8 1UG, United Kingdom }

\date{ }
\begin{document}
\maketitle
\begin{center}
francisco.glz.mty@gmail.com
\end{center}
\begin{center}
s.wiggins@bristol.ac.uk
\end{center}

\begin{abstract}

We study the phase space objects that control the transport in a classical Hamiltonian model for a chemical reaction. This model has been proposed to study the yield of products in an ultracold exothermic reaction. In this model, two features determine the evolution of the system: a Van der Waals force and a short-range force associated with the many-body interactions. In the previous work, small random periodic changes in the direction of the momentum were used to simulate the short-range many-body interactions. In the present work, random Gaussian bumps have been added to the Van der Waals potential energy to simulate the short-range effects between the particles in the system. We compare both variants of the model and explain their differences and similarities from a phase space perspective. To visualise the structures that direct the dynamics in the phase space, we construct a natural Lagrangian descriptor for Hamiltonian systems based on the Maupertuis action $S_0 = \int^{\mathbf{q}_f}_{\mathbf{q}_i} \mathbf{p} \cdot d\mathbf{q}$.

\end{abstract}

\newpage

\section{Introduction}
\label{sec:Intro}

Recent progress in experimental techniques allows the study of chemistry in certain cold systems \cite{Ospelkaus2010,Krems2005,Krems2008}. A current experiment of interest studies the products generated by the collision complex between two cold potassium-rubidium dimers,  

\begin{equation*}
2 \textrm{KRb} \rightarrow { [\textrm{K}_2 \textrm{Rb}_2] }^{\bigstar} \rightarrow \textrm{K}_2 + \textrm{Rb}_2.
\end{equation*}

Two ultra cold dimers $\textrm{KRb}$ meet at 300 nK, interact, and form a collision complex ${ [\textrm{K}_2 \textrm{Rb}_2] }^{\bigstar}$. The energy of the particles in this collision complex is around 4000 K. After this stage the final products, $\textrm{K}_2$ and $\textrm{Rb}_2$, are generated. The final energy of this products is expected to be around 14 K \cite{Ospelkaus2010,Ni2008,Hutson2010,Ni2010,Miranda2010,Marco2019}.

The considerably large difference of energy between the atoms in the collision complex and the final products is an important property to consider in the modelling of the system. A classical model has been proposed in \cite{Heller2018} to calculate the lifetime of the collision complex, or equivalently the yield of the final product in this cold chemical reaction. The justification for this model is based on semiclassical considerations. This model is an option to avoid the direct quantum calculations that require a large number of eigenstates to describe the dynamics of this kind of system with a deep potential well \cite{Bies2001}.

The basic idea in the construction of the model is that the long-range interaction in the system determines the escape of the particles when the energies are close to the threshold energy necessary to escape. For energies very close to the threshold, only the particles with enough momentum in the radial direction can escape from the potential well, which is referred to as the ``cauldron'' in \cite{Heller2018}. The short-range interactions are related to the collisions between molecules in the collision complex.

In the present work, we study the dynamics of this classical model from the phase space perspective. The phase space approach has been developed and applied recently to understand better the chemical reaction dynamics in different systems \cite{Waalkens2008, Teramoto2015, Tschope2020, Feldemair2019, Agaoglou2020, Garcia2020}. In Section \ref{sec:Model}, we describe the two variants of the model and their basic properties. Section \ref{sec:LD_S0} contains the construction of the Lagrangian descriptors based on the Maupertuis action $S_0$. In section \ref{sec:DPS}, we study the phase space of the integrable and the perturbed systems. Also, we compare the phase space structures involved in the dynamics of the trajectories that escape in the two variants of the model: the perturbed system and the kicked system. This study of the phase space helps us to understand the escape time of particles from the potential well in the in section \ref{sec:ET}. Finally, we present conclusions and remarks.

\newpage

\section{Model}
\label{sec:Model}

In this section, we study the basic features of the 2 degree of freedom classical model proposed in \cite{Heller2018} to estimate the yield of the final product in the chemical reaction described in section \ref{sec:Intro}. This model is inspired from previous work by Wannier \cite{Wannier1953} where the ionisation of electrons due to the collision between electrons in a helium atom is considered. With this model, Wannier obtained a threshold law for the yield of ionised electrons as a function of the energy. In both systems, the justification of the use of a classical approach is based on a semiclassical WKB analysis\cite{Wannier1953,Heller2018}.

An important characteristic of both models is that the total force in the system has two terms to define the escape process of the particles to infinity. The first force is a long-range interaction that determines the asymptotic motion. The second force has short-range interaction and in combination with the first force generates complicated dynamics in the region close to the origin.

\begin{figure}[!h]
\begin{center}
\includegraphics[scale=0.4]{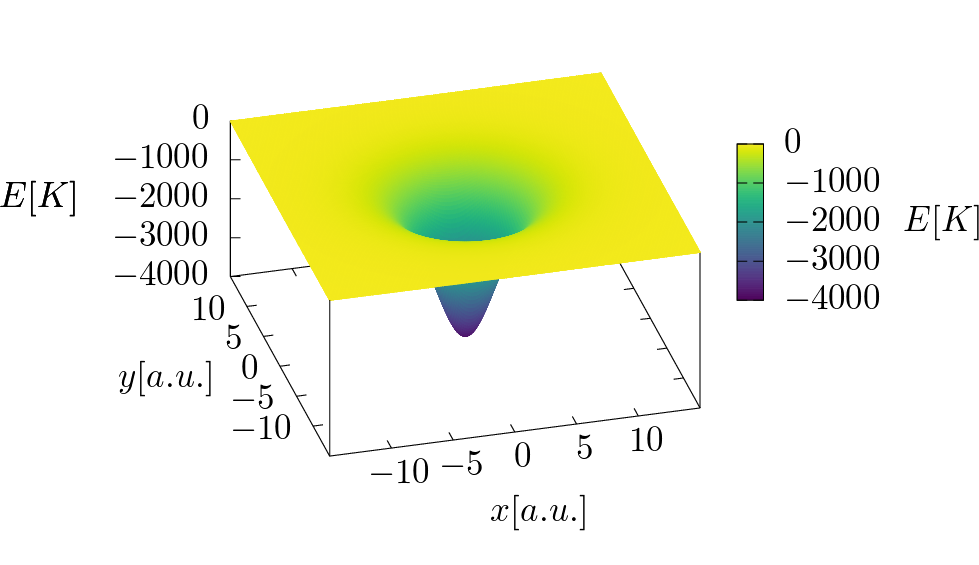}
\caption{Van der Waals potential energy $V_0(\mathbf{r})$. The energy $E$ is in Kelvin units. The cauldron is very deep, its minimum is around $-4000$ K. Similar potential energy surfaces have been proposed in \cite{Kraka2010,Montoya2020} for barrierless reactions.
\label{fig:Vo}}
\end{center}
\end{figure}

In the model for a cold chemical reaction, the system is considered as a two-body problem. The force related to the asymptotic motion is a Van der Waals force. The potential energy associated with this force is 

\begin{equation}
V_0(\mathbf{r}) = - \frac{C}{(\beta \left| \mathbf{r} \right| ^2 + \alpha )^3},
\label{eq:potencial_energy1}
\end{equation}

\noindent where $\mathbf{r}=(x,y)$ is the position from the origin, and the numerical value of the constants in this example are $C=16130$ a.u., $\beta= 2.9$ a.u., and $\alpha = 110$ a.u., see figure \ref{fig:Vo}. The potential energy $V_0(\mathbf{r})$ is negative and goes to zero when $|\mathbf{r}|$ goes to infinity. This potential energy has rotational symmetry and gives rise to integrable dynamics. For negative values of the total energy $E$, the phase space is bounded, and the particles are confined. For $E>0$ the phase space is unbounded, some particles can escape to infinity. 

Due to the radial symmetry of the potential $V_0(\mathbf{r})$, there exists an effective potential energy $V_{0eff}(r)$ parametrised by the $z$ component of the angular momentum, $L_z$. The threshold energy of the escape is determined by the maximum of the effective potential at the critical radius. Related to this radius exists a circular unstable periodic orbit that projects to a circle in the configuration space. This periodic orbit $\gamma_0$ is a normally hyperbolic invariant manifold (NHIM), almost any trajectory close to the orbit moves away from the orbit at an exponential rate after some time, only the trajectories in its stable manifold converge to the orbit $\gamma_0$. In the following sections, we explain the role of this family of hyperbolic periodic orbits in the dynamics of the system.

\begin{figure}
\begin{center}
\includegraphics[scale=0.4]{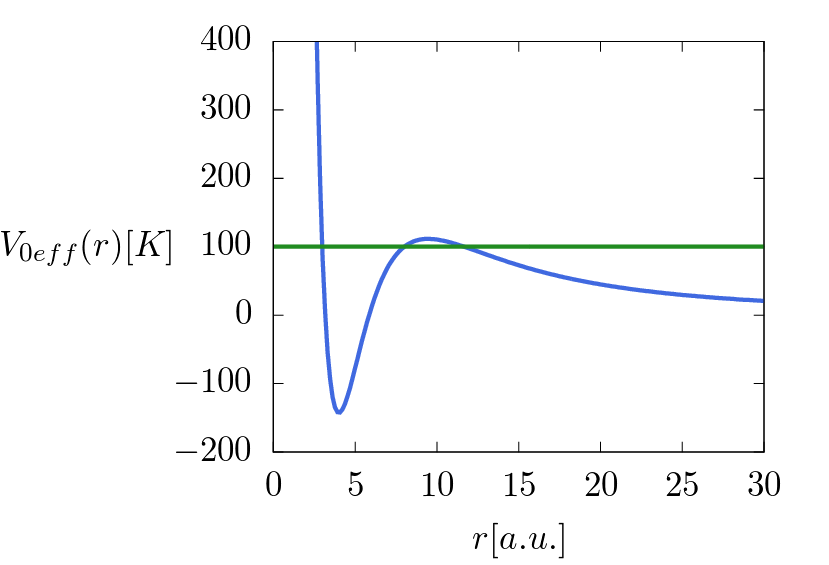}
\caption{Van der Waals effective potential energy $V_{0eff}(r)$. The maximum of $V_{0eff}(r)$ determines the radio of the hyperbolic periodic orbit $\gamma_0$. The numerical value of the $z$ component of the angular momentum is $L_z \simeq 108$ a.u. \label{fig:Veff}}
\end{center}
\end{figure}

The short-range force acts around the minimum of the potential $V_0(\mathbf{r})$ . In \cite{Heller2018} there are two proposals for the force to break the rotational symmetry and mimic the many-body interaction. The first proposal consists of adding to the potential $V_0(\mathbf{r})$ some random Gaussian bumps scattered inside the region with a radius $r<5$ a.u. . This kind of perturbation has been used in closed quantum systems to break their rotational symmetry and generate rich dynamics, and quasidegeneracy in the energy spectrum \cite{Luukko2016, Hellerbook}. In this case, the potential energy for this perturbed variant of the model is

\begin{equation}
V (\mathbf{r}) = V_0(\mathbf{r}) + \sum^{n}_{i = 1} A e^{ -B \left| \mathbf{r} - \mathbf{r_i} \right| ^2},
\label{eq:V}
\end{equation}

\noindent where $A=0.0001$ a.u., $B=10$ a.u.  are the coefficients that define the Gaussian bumps, and $r_i$ are the position of their centres. The figure shows a plot of the potential energy $V(\mathbf{r})$ in colour scale.

\begin{figure}[!h]
\begin{center}
\includegraphics[scale=0.4]{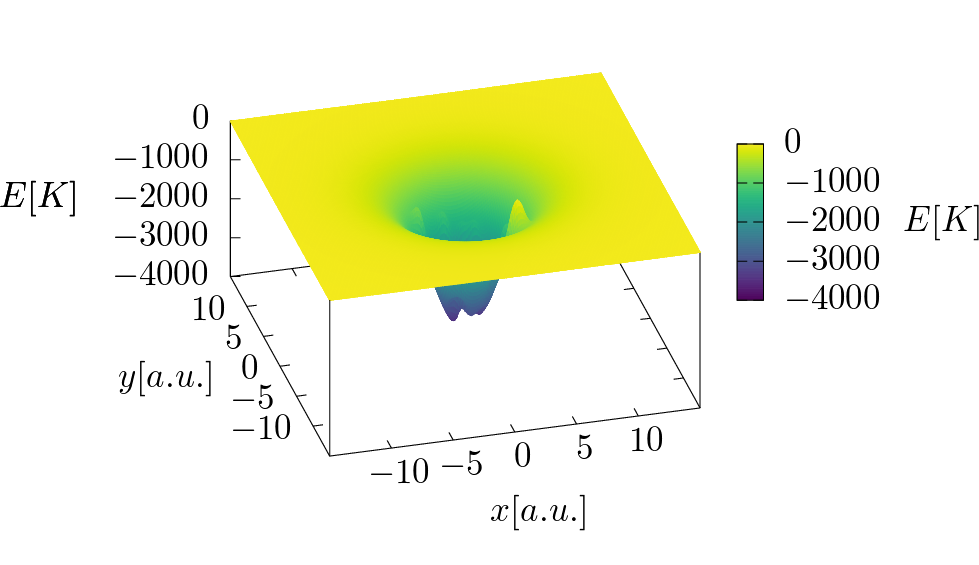}
\caption{Potential energy for the perturbed system $V(\mathbf{r})$. The values of $A=0.002$ a.u. and $B=10$ a.u. have been chosen in this plot so that the random bumps appreciably break the symmetry. For the numerical calculations we used $A=0.0001$ a.u. and $B=10$ a.u.
\label{fig:V}}
\end{center}
\end{figure}

The second variant of the short-range force has been proposed to simplify the numerical calculations of the trajectories. The basic idea is to generate trajectories ``similar'' to the trajectories for the perturbed system without the inclusion of the random bumps in the potential energy that generate instabilities in the numerical calculations. The alternative explored in \cite{Heller2018} is the use of small periodic random changes in the direction of the momentum of the particles. These changes in the momentum are only possible if the particles are in the same region where the bumps are in the other variant of the model, $r<5$ a.u. These changes in the direction of momentum preserve the energy of the particle and the numerical calculations of the trajectories are more simple than the calculations for the trajectories under the perturbed potential $V(\mathbf{r})$. This method is called random momentum kicks.  

\begin{figure}[h!]
\begin{center}
 \subfigure[Integrable trajectory.]{ 
 \includegraphics[scale=0.25]{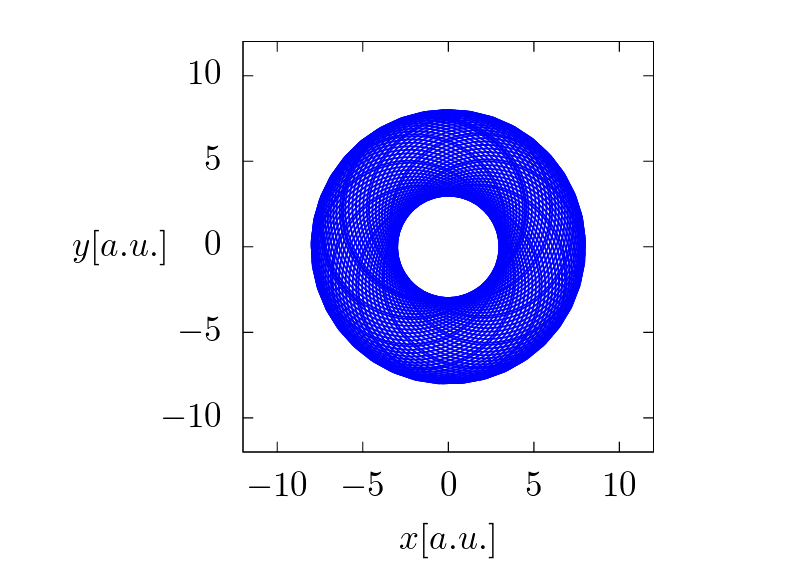}}
 \subfigure[ $\frac{\Delta E}{E}$ for the integrable trajectory.]{
\includegraphics[scale=0.25]{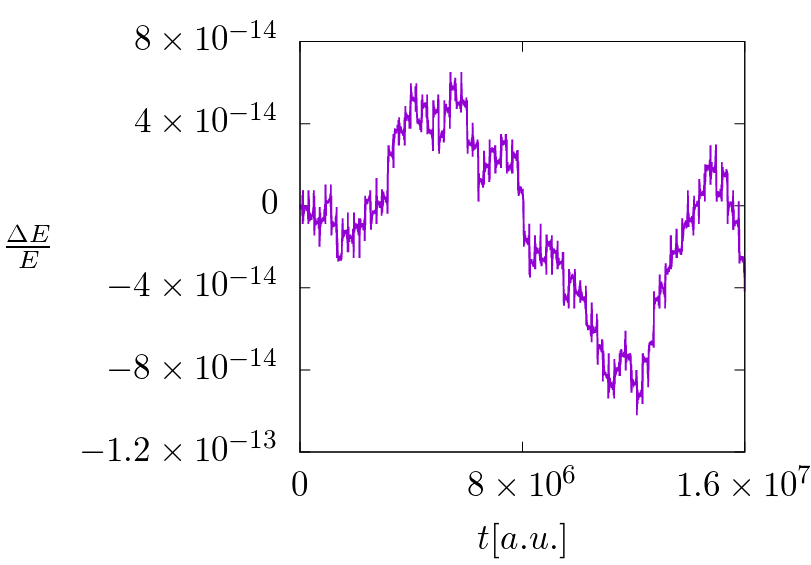}}\\
\subfigure[Nonintegrable trajectory.]{
\label{fig:trajectories_energies_c}
\includegraphics[scale=0.25]{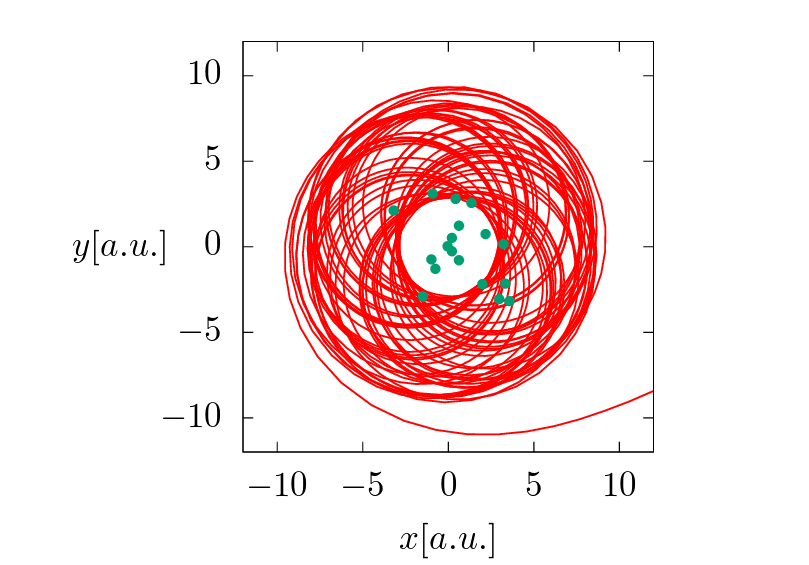}}
\subfigure[ $\frac{\Delta E}{E}$ for the nonintegrable trajectory.]{
\includegraphics[scale=0.25]{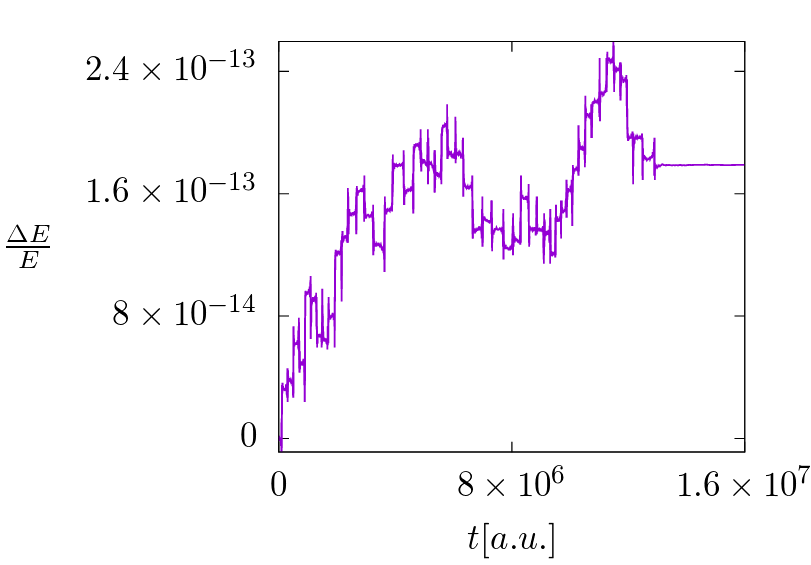}} \\
\subfigure[Kicked trajectory.]{ 
\label{fig:trajectories_energies_e}
\includegraphics[scale=0.25]{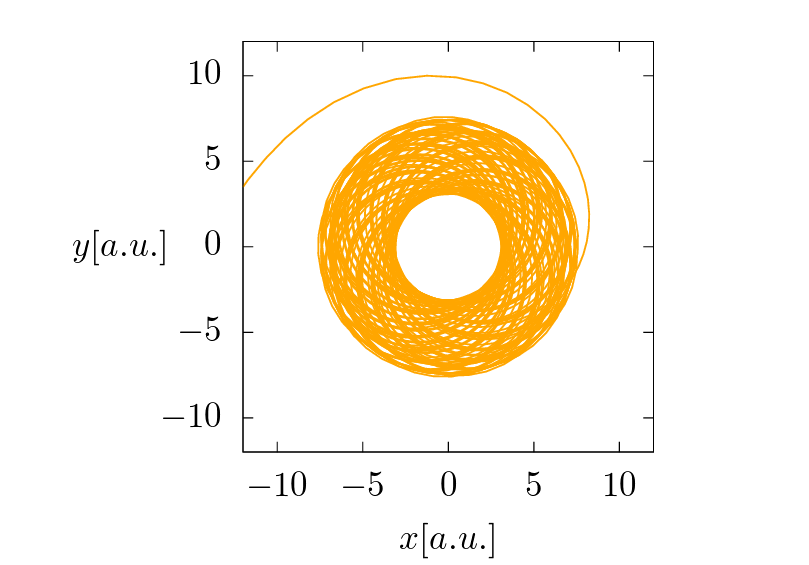}}
 \subfigure[ $\frac{\Delta E}{E}$ for the kicked trajectory.]{
\includegraphics[scale=0.25]{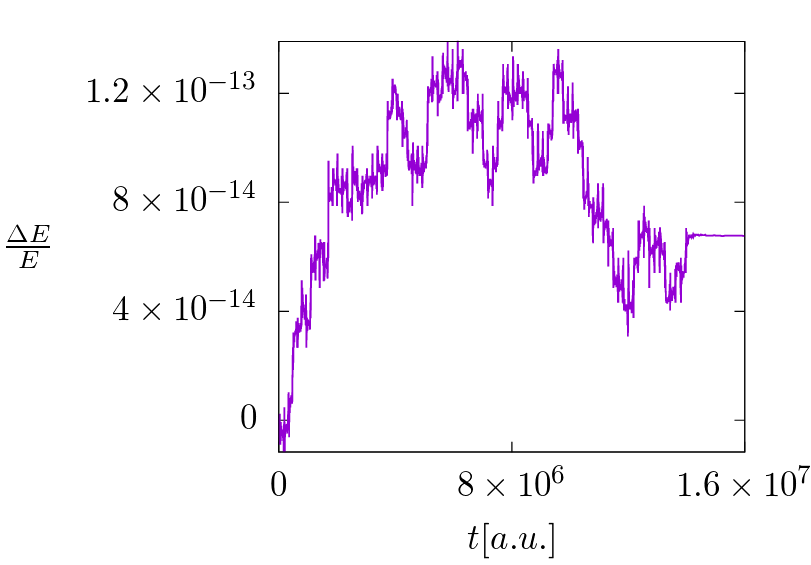}}
\caption{On the left side, the trajectories in the configuration space for the three considered cases: integrable, perturbed nonintegrable, and kicked. On the right side, their corresponding relative changes of the energy $\frac{\Delta E}{E}$. The time $t$ is in atomic units a.u. Their initial energies angular momentum are equal in the three cases, $E = 100$ K and $L_z\simeq 108$ a.u. The green dots on the plot of the trajectory corresponding to the perturbed system indicate the centres of the Gaussian bumps in the potential $V(\mathbf{r})$. For these values of $E$ and $L_z$ the particle can not escape in the integrable case, see $V_{0eff}(r)$ in figure \ref{fig:Veff}. However, the short-range interactions present in the nonintegrable and kicked system change the dynamics, and the particle can escape to infinity. 
\label{fig:trajectories_energies}}
\end{center}
\end{figure}

The figure \ref{fig:trajectories_energies} shows the trajectories with the same initial conditions for the three cases and its corresponding relative changes in the energy. The numerical calculations of the trajectories are done with a Taylor polynomial integrator order 21 implemented in the language Julia \cite{Perez2019, Benet2018, Benet2019}. Using this integrator, the relative changes of the energies in the three cases have a similar order of magnitude. The blue trajectory corresponds to a particle under the influence of the Van der Waals potential $V_0(\mathbf{r})$. This trajectory is bounded and quasiperiodic. The red trajectory corresponds to a particle under the perturbed potential $V(\mathbf{r})$. This red trajectory has a more complicated behaviour generated by the Gaussian bumps before to escape to the asymptotic region. The orange trajectory is obtained with the random kicks method, after some time close to the origin the particle escapes to infinity.

\newpage

\newpage

\section{Lagrangian Descriptors and the Maupertuis action $S_0$}
\label{sec:LD_S0}

The classical Lagrangian descriptor is based on the arc length of the trajectories in the phase space \cite{madrid2009distinguished, lopesino2017}. In the present article, we construct a Lagrangian descriptor based on the classical action $S_0$ to reveal the phase space structures. The action $S_0$ is an essential quantity in the study of Hamiltonian systems and plays a fundamental role in semiclassical approximations. In this section, we explain why it is possible to construct a Lagrangian descriptor based on $S_0$. First, let us consider the definition of the Lagrangian descriptor proposed in \cite{lopesino2017} and basic ideas about the phase space structures that determine the dynamics. 

The general definition of the Lagrangian descriptor is as follows. Consider a system of ordinary differential equations

\begin{equation}
\frac{d \mathbf{x}}{dt} = \mathbf{v}(\mathbf{x}), \quad \mathbf{x} \in \mathbb{R}^n \;,\; t \in \mathbb{R},
\end{equation}

\noindent where $\mathbf{v}(\mathbf{x}) \in C^r$ ($r \geq 1$) in $\mathbf{x}$ and continuous in time $t$. The definition of Lagrangian descriptor depends on the initial condition $\mathbf{x}_{0} = \mathbf{x}(t_0)$, on the time interval $[t_0+\tau_{-},t_0+\tau_{+}]$, and takes the form,

\begin{eqnarray}
M(\mathbf{x}_{0},t_{0},\tau_{+},\tau_{-}) & = & M_{+}(\mathbf{x}_{0},t_{0},\tau_{+}) + M_{-}(\mathbf{x}_{0},t_{0},\tau_{-}) \nonumber \\
& = & 
\displaystyle{ \int^{t_{0}+\tau_{+}}_{t_{0}} F(\mathbf{x}(t))\; dt + \int^{t_{0}}_{t_{0}+\tau_{-}} F(\mathbf{x}(t)) \; dt ,} \label{eq:LD}
\end{eqnarray}

\noindent where a $F$ is a positive function on the solutions $\mathbf{x}(t)$, $ \mathbf{x}(t_0) = \mathbf{x}_{0}$ and $\tau_{+} \geqslant 0$ and $ \tau_{-} \leqslant 0 $ are freely chosen parameters. The values of $\tau_{+}$ and $\tau_{-}$ can change between different initial conditions and allow us to stop the integration once a trajectory leaves a specific region in the phase space. In this way it is possible to reveal only the phase space structures contained in a defined region.

The phase space of a 2 degree of freedom Hamiltonian system has 4 dimensions. Considering the conservation of energy, we can represent the dynamics of the system in the 3-dimensional constant energy level set. In this 3-dimensional manifold, we can visualise the dynamics and identify the essential phase space structures that direct the dynamics.

The periodic orbits are basic objects for understanding the dynamics of the system in the constant energy level set. Around a stable periodic orbit, the KAM-tori confine the trajectories in a bounded region defined by the tori. In contrast, the dynamics in a neighbourhood of an unstable hyperbolic periodic orbit has different behaviour, the trajectories around the orbit diverge from the orbit after some time. There are two invariant surfaces under the flow associated with a hyperbolic periodic orbit, the stable and unstable manifold of the hyperbolic periodic orbit.
Its stable and unstable manifolds intersect in the hyperbolic periodic orbit and direct the trajectories in the neighbourhood. The definition of the stable and unstable manifolds $W^{s/u} (\gamma)$ of the hyperbolic periodic orbit $\gamma$ is the following,

\begin{equation}
 W^{s/u} (\gamma) = \lbrace \mathbf{x} \vert \mathbf{x}(t) \rightarrow \gamma, t \rightarrow \pm \infty \rbrace.
\end{equation}

The stable manifold $W^s(\gamma)$ is the set of trajectories that converge to the periodic orbit $\gamma$ as the time $t$ goes to $\infty$. The definition of the unstable manifold $W^u(\gamma)$ is similar. The unstable manifold is the set of trajectories converging to the periodic orbit as the time $t$ goes to $-\infty$. 

In a 2 degree of freedom Hamiltonian system, the invariant manifolds $W^{s/u} (\gamma)$ are 2-dimensional surfaces. These surfaces form impenetrable barriers that direct the dynamics in the 3-dimensional constant energy level set \cite{Kovacks2001,Wiggins2001}. Another important property of the stable and unstable manifolds related to the chaotic dynamics is that, if a stable manifold and an unstable manifold intersect transversally at one place, then there are an infinite number of transversal intersections between them. The structure generated by the union of the stable and unstable manifolds is called a tangle. It defines a set of tubes that direct the dynamics in the constant energy level set \cite{ de1990geometry, de1991cylindrical}. A remarkable property of the dynamics in the constant energy level set is that the trajectories in a tube never cross the boundaries of a tube. This fact is a consequence of the uniqueness of the solutions of the ordinary differential equations. 

The Lagrangian descriptors are appropriate tools to reveal the phase space structure, especially, to find stable and unstable manifolds of periodic orbits \cite{ demian2017detection, naik2019finding, naik2020detecting}. To understand the basic idea that holds up the detection, let us consider the behaviour of the trajectories in a neighbourhood of a stable manifold $W^{s}(\gamma)$. All the trajectories in $W^{s} (\gamma)$ converge to the periodic orbit $\gamma$, and the trajectories in a small neighbourhood of $W^{s} (\gamma)$ have similar behaviour just for a finite interval of time. After this interval of time, the trajectories move apart from the unstable hyperbolic periodic orbit $\gamma$ following the unstable manifold $W^u(\gamma)$. This different behaviour of the trajectories generates the singularities in the Lagrangian descriptors, and other chaotic indicators, like scattering functions \cite{Gonzalez2012,Drotos2014,Gonzalez2020}.

The stationary action principle developed by Leibniz, Euler, and Maupertuis establishes that the action $S_0$ of a Hamiltonian system defined as 

\begin{equation}
S_0 = \int^{\mathbf{q}_f}_{\mathbf{q}_i} \mathbf{p} \cdot d\mathbf{q} 
\end{equation} 

\noindent has an extreme value on the trajectory of the system. The quantities $\mathbf{p}$ and $\mathbf{q}$ are the generalised momenta the generalised coordinates of the system. It is possible to construct a natural Lagrangian descriptor for Hamiltonian systems under the following considerations. 

Let us consider a system such that the kinetic energy is a quadratic function of the generalised velocities $\dot{\mathbf{q}}$. Then

\begin{equation}
T = \frac{d\mathbf{q}}{dt} \; \mathbf{M(q)} \; \frac{d\mathbf{q^\intercal }}{dt},
\end{equation}

\noindent where $\mathbf{M(q)}$ is the mass tensor and it is a function only of the generalised coordinates $\mathbf{q}$. For such systems, exist an identity between the kinetic energy, the generalised momenta, and the generalised velocities, 

\begin{equation}
2T = \mathbf{p} \cdot \dot{\mathbf{q}},
\end{equation}

\noindent provided that the potential energy $V({\mathbf {q}})$ is not a function of $\dot{\mathbf{q}}$. By defining a distance $ds$ in the space of generalised coordinates

\begin{equation}
 ds^2 = d\mathbf{q} \; \mathbf{M(q)} \; d \mathbf{q^\intercal}, 
\end{equation}

\noindent one recognises the mass tensor $\mathbf{M(q)}$ as a metric tensor. The kinetic energy can be written as 

\begin{equation}
T = \frac{1}{2} \left( \frac{ds}{dt} \right)^2
\end{equation}

\noindent or, equivalently, 

\begin{equation}
2Tdt = \mathbf{p} \cdot d \mathbf{q} = \sqrt{2T} \; ds.
\end{equation}

Hence, the action $S_0$ can be expressed as

\begin{equation}
S_0 = \int^{\mathbf{q}_f}_{\mathbf{q}_i} \mathbf{p} \cdot d\mathbf{q} = 
\int^{\mathbf{q}_f}_{\mathbf{q}_i} \sqrt{2 (E- V(\mathbf{q}))} \; ds = 2 \int^{t_f}_{t_i} T \; dt.
\label{eq:S_K}
\end{equation}

\noindent Therefore, the quantity $\mathbf{p} \cdot d \mathbf{q}$ and its integral, the action $S_0$, are positive quantities along any trajectory in the phase space and can, therefore, be used to construct a Lagrangian descriptor to study the phase space for this type of Hamiltonian systems.

The Lagrangian descriptor $M_{S_{0}}$ based on the action $S_0$ evaluated at
times $\tau_{-}$, $\tau_{+}$ and the point $ \mathbf{x}_{0} = \mathbf{x}( t_{0} ) = ( \mathbf{q}_{0},\mathbf{p}_{0})$ on the trajectory $\mathbf{x}(t)=(\mathbf{q}(t),\mathbf{p}(t))$ is defined as

\begin{eqnarray}
M_{S_{0}}(\mathbf{x}_{0},t_{0},\tau_{+},\tau_{-}) & = & S_{0_+}(\mathbf{x}_{0},t_{0},\tau_{+}) + S_{0_-} (\mathbf{x}_{0},t_{0},\tau_{-}) \nonumber \\
& = & 
\displaystyle{ \int^{ \mathbf{q}_{+} }_{ \mathbf{q}_{0} } \mathbf{p} \cdot d\mathbf{q} + \int^{ \mathbf{q}_{0} }_{ \mathbf{q}_{-} } \mathbf{p} \cdot d\mathbf{q} }\nonumber \\
& = & 
\displaystyle{ 2 \int^{t_{0}+\tau_{+}}_{t_{0}}  T(\mathbf{x}(t))\; dt + 2 \int^{t_{0}}_{t_{0}+\tau_{-}}  T(\mathbf{x}(t)) \; dt .} \label{eq:LD_S}
\end{eqnarray}

\newpage

\section{Dynamics and phase space}
\label{sec:DPS}

In order to understand the dynamics of the perturbed and kicked systems, it is convenient to begin by analysing the phase space of the integrable system. The phase space structures in the integrable system are easy to visualise and serve as a reference to study the structures in the other two cases. We use the Lagrangian descriptor based on the action $M_{S_{0}}$, constructed in the section \ref{sec:LD_S0}, and the Poincare map as tools to visualise the relevant structures in the phase space to understand the dynamics.

\subsection{Phase space of the integrable system}

\begin{figure}[h!]
\begin{center}
 \subfigure[$E=0.1$ K]{ 
\includegraphics[scale=0.25]{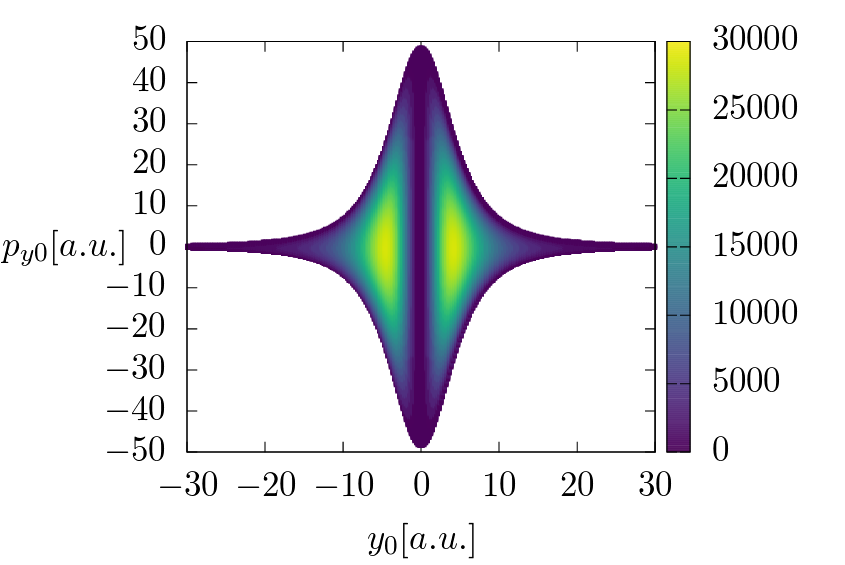}}
 \subfigure[$E=1$ K]{
\includegraphics[scale=0.25]{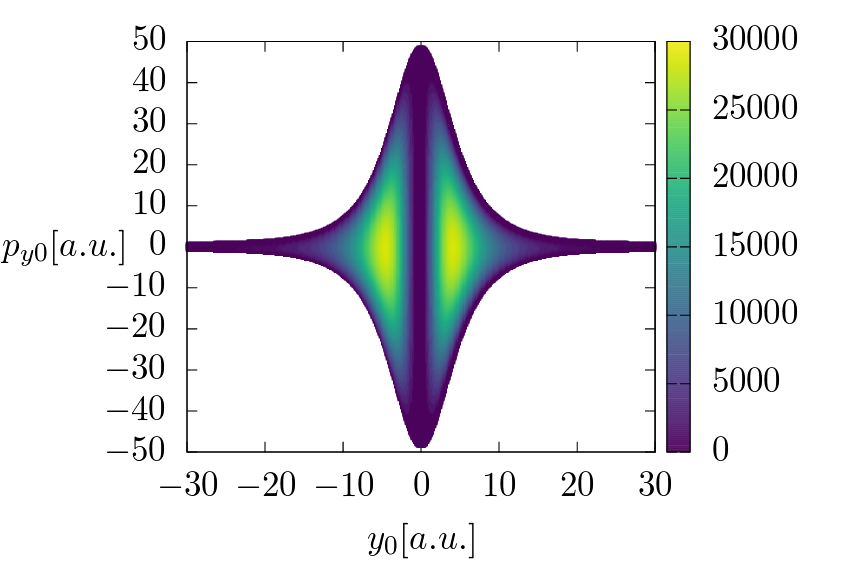}}\\
\subfigure[$E=10$ K]{
\includegraphics[scale=0.25]{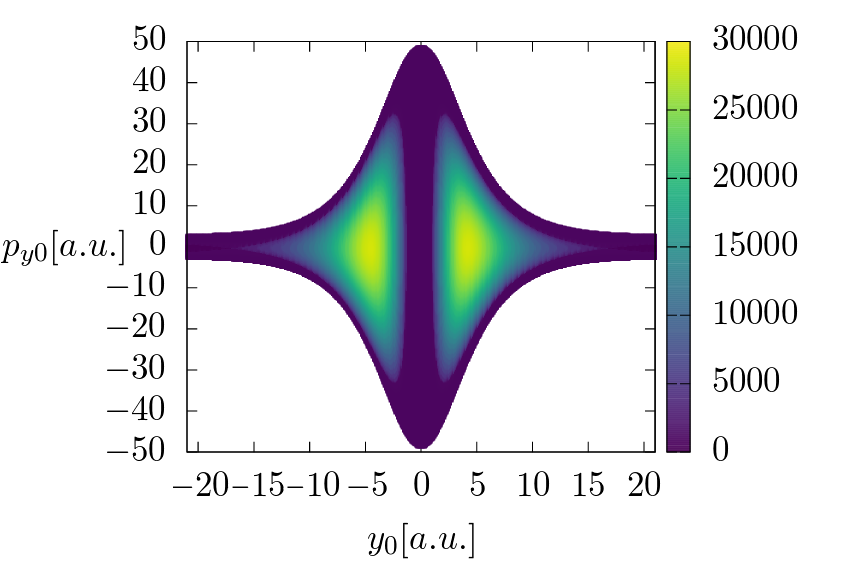}}
\subfigure[$E=100$ K]{
 \label{fig:LD_integrable_energies_d}

\includegraphics[scale=0.25]{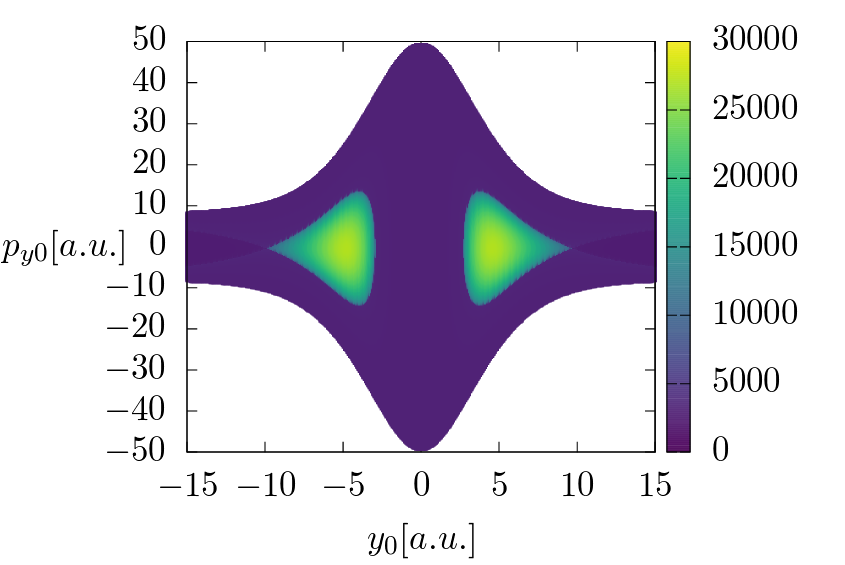}} 

\caption{ Lagrangian descriptor $M_{S_{0}}$ with initial conditions on the plane $y_0$--$p_{y0}$ at $x_0=0$, and $p_{x0}>0$ for different values of the energy $E$. The value of the integration times are $\tau_+,-\tau_- = 2 \times 10^6 $ a.u. \label{fig:LD_integrable_energies}}

\end{center}
\end{figure}

The integrable system has rotational symmetry. Therefore a natural choice of initial conditions to analyse the dynamics is a set that considers this symmetry. The figure \ref{fig:LD_integrable_energies} shows the Lagrangian descriptor $M_{S_0}$ evaluated in the canonical conjugate plane $y_0$--$p_{y0}$ at $x=0$ and $p_{x0}>0$ compatible with the conservation of the energy. In the Lagrangian descriptor plots exist two bounded regions with a large value of Lagrangian descriptor (yellow-green regions) and an unbounded region around with low values of the Lagrangian descriptor (dark-blue region). 

The region dark-blue colour with small values of Lagrangian descriptor corresponds to trajectories that escape to infinity and the regions with large values of Lagrangian descriptor in yellow-green correspond to trapped trajectories. To explain this property in the Lagrangian descriptor plots, let us consider the equation \ref{eq:LD_S}, the conservation of the energy, and only the positive time direction. Analogous considerations follow for the negative time direction. For a fixed value of the energy $E>0$, the values of the kinetic energy are smaller in the asymptotic region than close to the minimum of $V_0(\mathbf{r})$. The kinetic energy of the particles that escape to infinity converges to its minimum possible value, the total energy $E$, see figure \ref{fig:Vo}. The integral of the kinetic energy with respect to the time is proportional to the action $S_0$. Then, for any large enough finite interval of time, the Lagrangian descriptor is smaller for unbounded trajectories than for the trapped trajectories. The trapped trajectories form integrable islands around the stable periodic orbit corresponding to the minimum of the $V_{0eff}(r)$. 

In the integrable case, the boundary of the stable islands is defined by the stable and unstable manifolds of the hyperbolic periodic orbit $\gamma_0$ associated with the maximum in the effective potential $V_{0eff}(r)$. The hyperbolic periodic orbit $\gamma_0$ corresponds to the point with $p_{y0} = 0 $ and the maximal values of $y_{0}$ in the green-yellow region. The symmetric point with respect to the $p_{y0}$ axis corresponds to the analogous periodic orbit with $-L_z$, for simplicity we consider only the orbit $\gamma_0$ in the following argumentation. At this point, two lines that extend to the asymptotic region intersect, see figures \ref{fig:LD_integrable_energies_d} and \ref{fig:poincare1_kicked} corresponding to $E=100$ K. Those lines, where the value of the Lagrangian descriptor have a sharp peak, are the intersections of the stable and unstable manifolds $W^{s/u}(\gamma_0)$ with the set of initial conditions. These invariant manifolds have dimension 2 and divide the constant energy level set.

\subsection{Phase space of the perturbed system}

\begin{figure}[h!]
\begin{center}
 \subfigure[$E=0.1$ K]{ 
\includegraphics[scale=0.25]{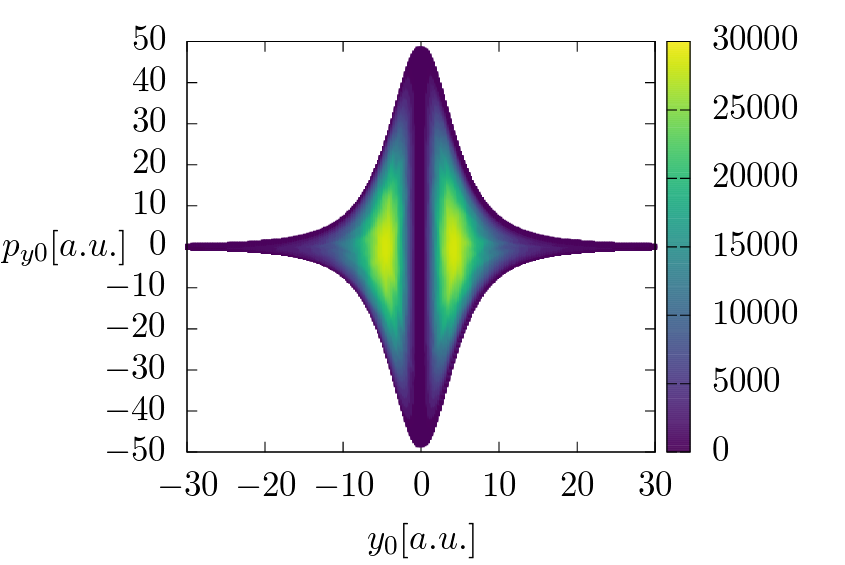}}
 \subfigure[$E=1$ K]{
\includegraphics[scale=0.25]{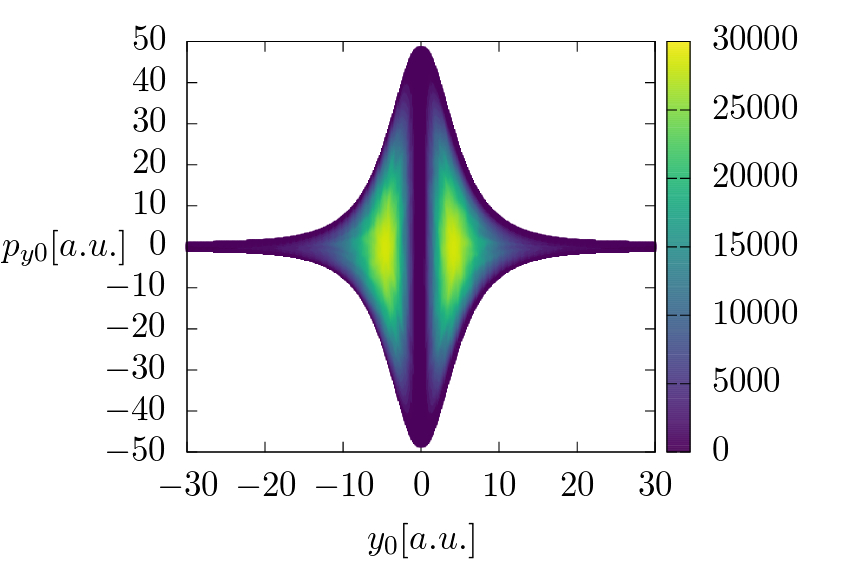}}\\
\subfigure[$E=10$ K]{
\includegraphics[scale=0.25]{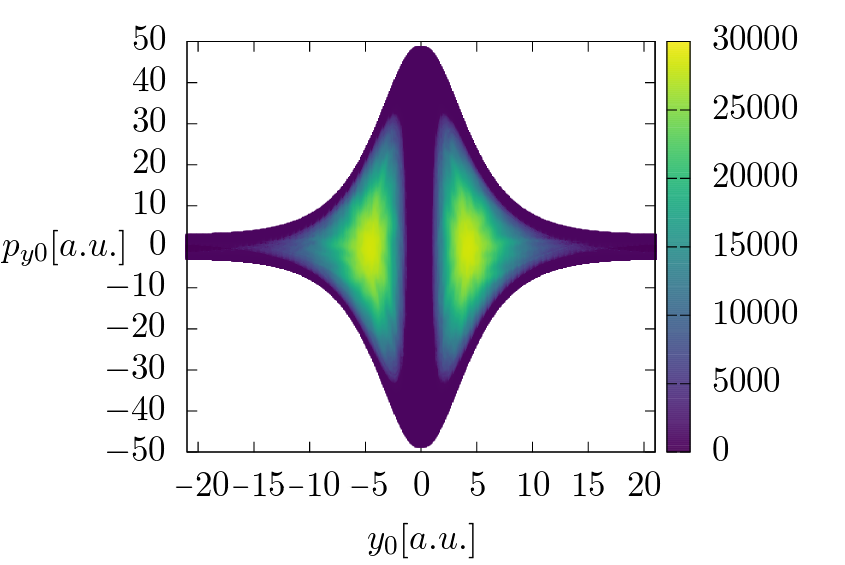}}
\subfigure[$E=100$ K]{
\includegraphics[scale=0.25]{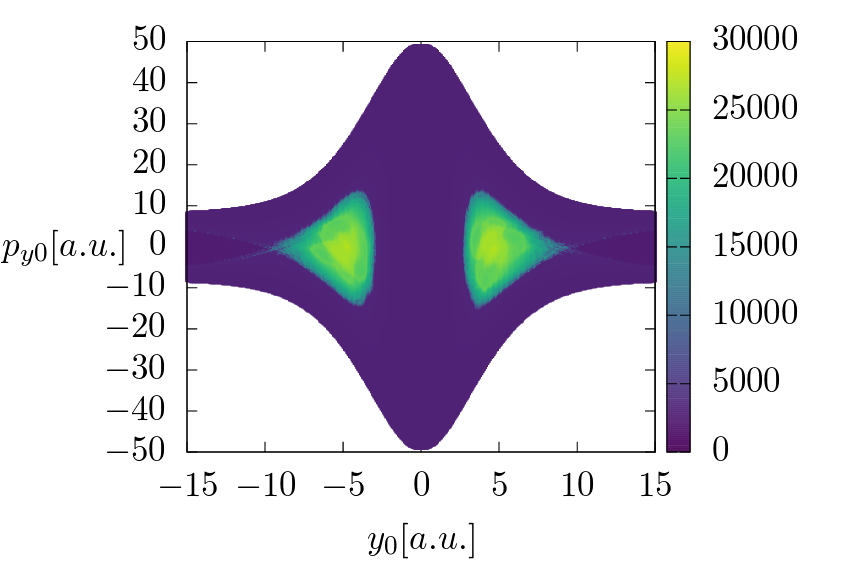}} 

\caption{Lagrangian descriptor $M_{S_{0}}$ with initial conditions on the plane $y_0$--$p_{y0}$ and $p_{x0}>0$ for different values of the energy $E$. The value of the integration times are $\tau_+,-\tau_- = 2 \times 10^6 $ a.u. \label{fig:LD_nonintegrable_energies}}
\end{center}
\end{figure}

Next, we analyse the phase space of the perturbed system defined by the potential energy $V(\mathbf{r})$. In order to compare the results with the integrable case, it is convenient to consider the same kind of initial conditions. In the figure \ref{fig:LD_nonintegrable_energies} there are plots of the Lagrangian descriptor evaluated in the plane $y_0$--$p_{y0}$ at $x=0$ and $p_{x0}>0$ compatible with the conservation of the energy. The values of the energies are the same as those chosen in the integrable case.

The plots of the Lagrangian descriptors in the figure \ref{fig:LD_nonintegrable_energies} shows changes with respect to the plots in the figure \ref{fig:LD_integrable_energies} corresponding to the integrable case. The values of the Lagrangian descriptor are more irregular in the regions with a large value of Lagrangian descriptors as a result of the random Gaussian bumps in the potential energy $V(\mathbf{r})$. However, the blue regions associated with the trajectories that escape to the asymptotic region are similar in both cases.

\newpage

\begin{figure}[H]
\begin{center}
 \subfigure[$E=0.1$ K]{ 
\includegraphics[scale=0.25]{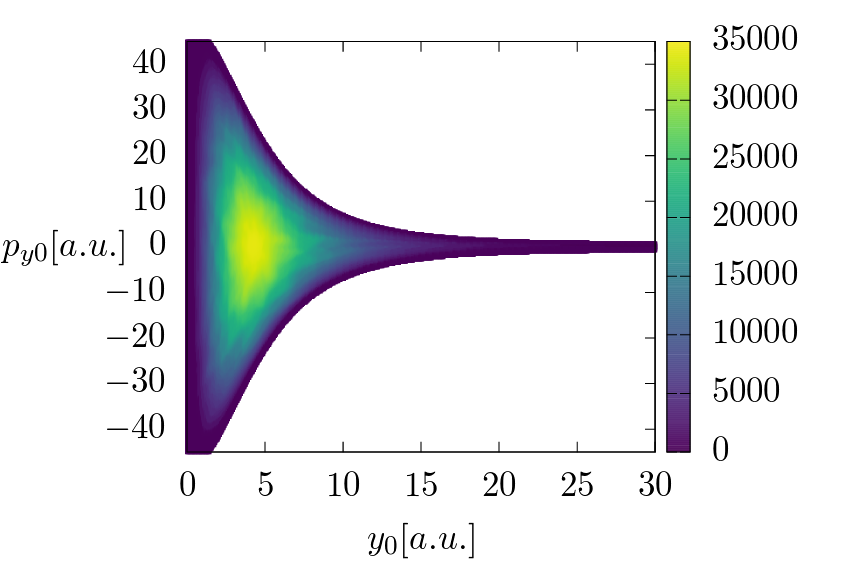}}
 \subfigure[$E=0.1$ K]{ 
\includegraphics[scale=0.25]{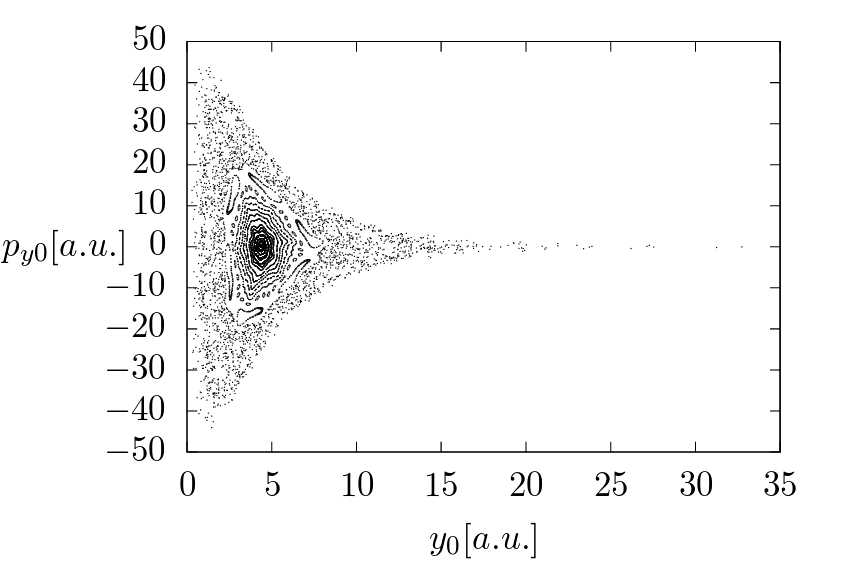}}\\
 \subfigure[$E=1$ K]{
\includegraphics[scale=0.25]{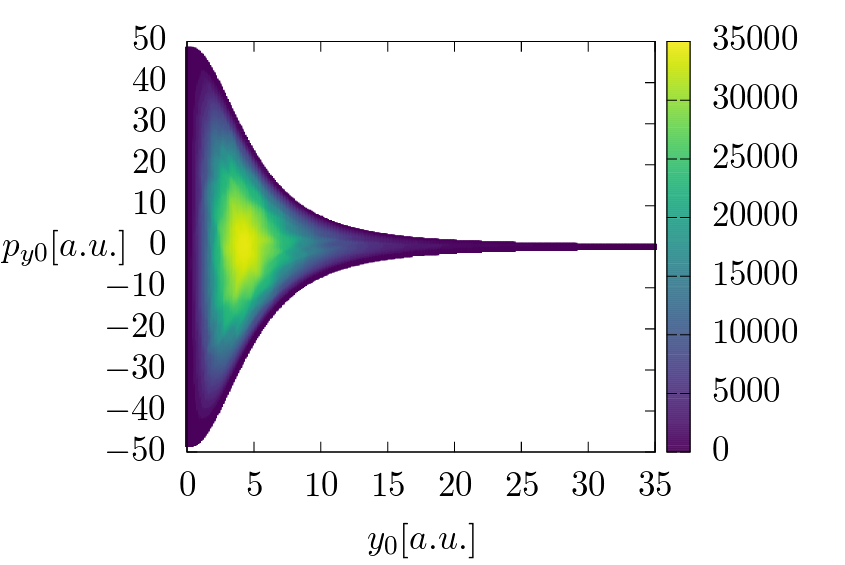}}
 \subfigure[$E=1$ K]{
\includegraphics[scale=0.25]{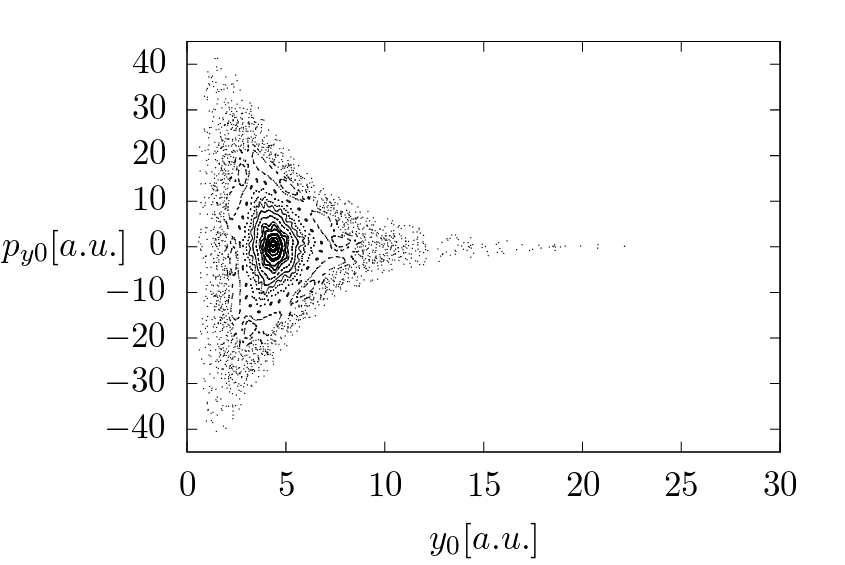}}\\
\subfigure[$E=10$ K]{
\includegraphics[scale=0.25]{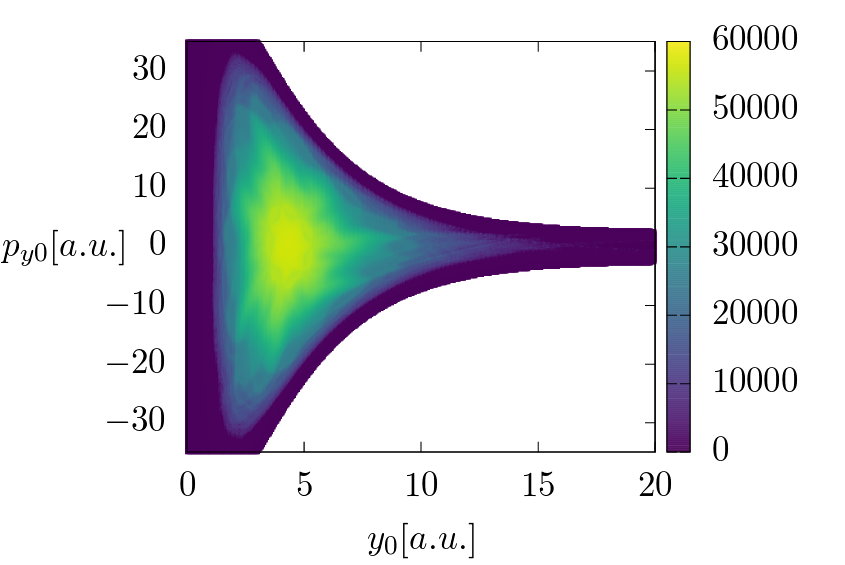}}
\subfigure[$E=10$ K]{
\includegraphics[scale=0.25]{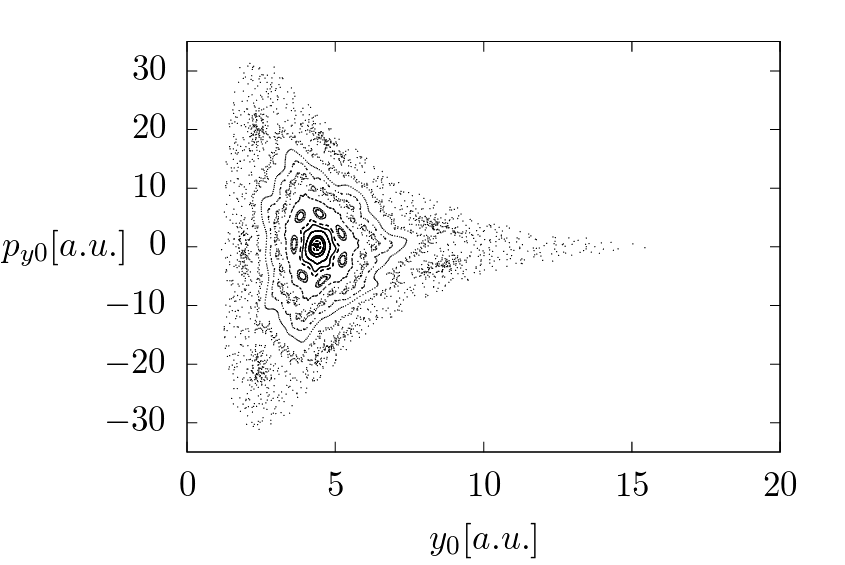}}\\
\subfigure[$E=100$ K]{
 \label{fig:LD_poincare_nonintegrable_energies_g}
\includegraphics[scale=0.25]{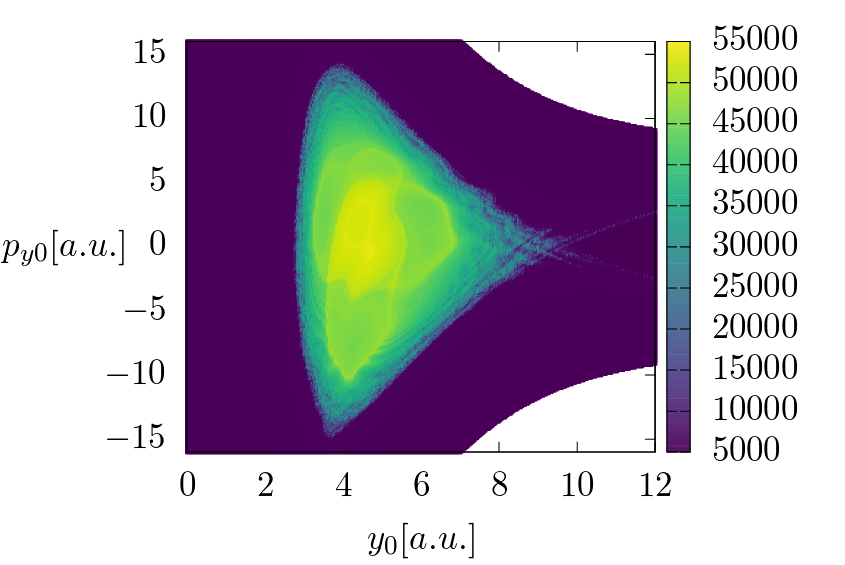}} 
\subfigure[$E=100$ K]{
\includegraphics[scale=0.25]{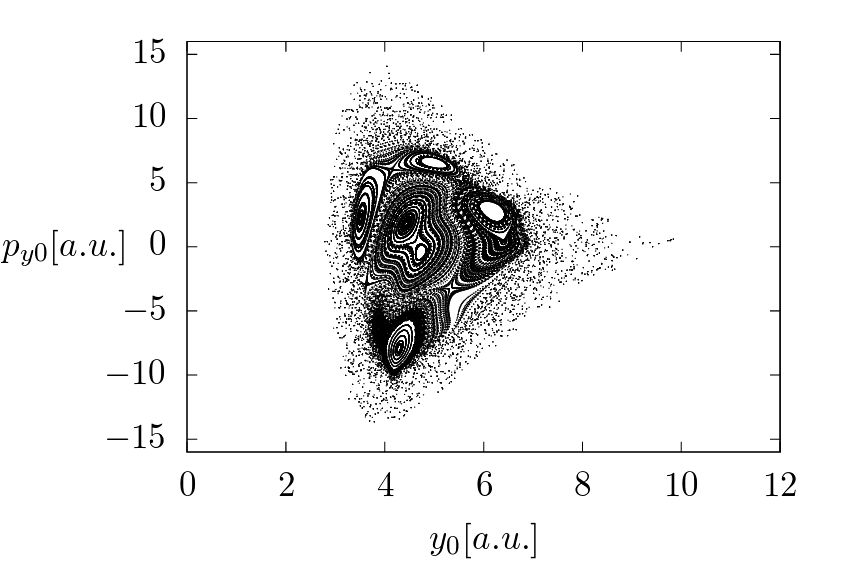}}

\caption{ Magnification of the Lagrangian descriptor $M_{S_{0}}$ plots for the nonintegrable system in the figure \ref{fig:LD_nonintegrable_energies} and their corresponding Poincare maps. The value of the integration times for the plots are $\tau_+,-\tau_- = 1 \times 10^6 $ a.u. Plots on the right side show only the intersections of the trajectories with the Poincare plane that remain in the domain after some time. In this manner, it is easy to distinguish the regions that escape fast (the external large white region), the transient chaotic sea generated by the homoclinic tangle of $\gamma$, (the region where the intersections form an irregular pattern), and the stable KAM islands (the region where the iteration form closed curves). \label{fig:LD_poincare_nonintegrable_energies}}
\end{center}
\end{figure}

\begin{figure}[H]
\begin{center}
\includegraphics[scale=0.4]{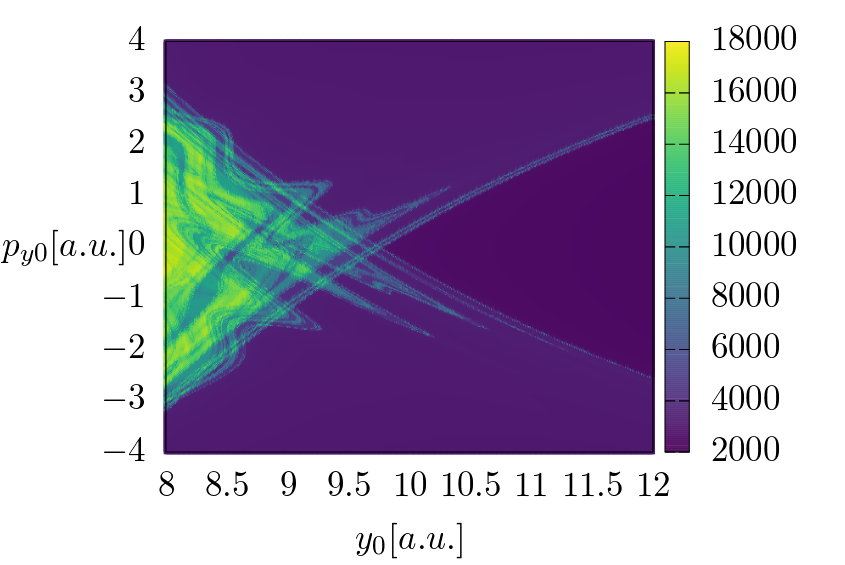}
\caption{Magnification of the Lagrangian descriptor $M_{S_{0}}$ plot for the nonintegrable system for $E = 100$ K in figure \ref{fig:LD_poincare_nonintegrable_energies_g}. The periodic orbit $\gamma$ intersects the plane at the corner of the blue triangle with low values of $M_{S_{0}}$, close to the point $(10,0)$. This periodic orbit is near to its corresponding hyperbolic orbit $\gamma_0$ in the integrable system. The Gaussian perturbations decay very fast, and its contribution to the potential energy $V(\mathbf{r})$ is small in the neighbourhood of the periodic orbit $\gamma_0$. However, the stable and unstable manifolds of $\gamma$ intersect transversely and form a chaotic homoclinic tangle. These manifolds determine the entry and exit from the region around the KAM islands. The size of the exit lobes is small compared with the transient chaotic sea around the KAM islands. The value of the integration times for this plot are $\tau_+,-\tau_- = 2 \times 10^6 $ a.u.\label{fig:zoom_zoom} }
\end{center}
\end{figure}

To appreciate better the details in the Lagrangian descriptor for the nonintegrable case, the figure \ref{fig:LD_poincare_nonintegrable_energies} shows magnifications of the region with $y_{0}>0$ and its corresponding Poincare maps. The trajectories to construct Poincare maps have been trapped in this region for some time, and only their intersections after this time are plotted. In the Poincare maps, there are some KAM islands. After some time the trajectories outside the KAM islands escape to the asymptotic region where the motion is simple. The KAM islands are surrounded by trajectories with a temporal irregular behaviour determined by the tangle between the stable and unstable manifolds of the external hyperbolic periodic orbit $\gamma$, see Lagrangian descriptor plot in figure \ref{fig:zoom_zoom}. The orbit $\gamma$ is the deformation of the original hyperbolic periodic orbit $\gamma_0$ generated by the perturbation. 

The irregular temporal behaviour of the trajectories around the KAM islands is an example of a phenomenon called transient chaos \cite{Tel2015,Telbook,Janosi2019}. This complicated transient behaviour is common in open Hamiltonian systems. Some recent studies of the phase space structures of open Hamiltonian systems with two and three degrees of freedom are in\cite{GC2011,Sanjuan2013,Gonzalez2012,Drotos2014,Gonzalez2020}.

\newpage

\newpage

\newpage

\subsection{Comparison between the dynamics of the perturbed nonintegrable system and the kicked system}

In the nonintegrable system, the Gaussian perturbations change the dynamics around the origin. The KAM islands are surrounded by the transient chaotic sea generated by the homoclinic tangle of the periodic orbit $\gamma$. To appreciate more details about the dynamics around the KAM islands, we consider the Poincare map of red trajectory in figure \ref{fig:trajectories_energies_c}. The Poincare map and the Lagrangian descriptor $M_{S_0}$ as a background are in figure \ref{fig:LD_poincare_nonintegrable_energies}. The Lagrangian descriptor reveals the complicated structure of the tangle between the stable and unstable manifolds $W^{s/u}(\gamma)$. The size of the lobes where the trajectories escape to infinity is small compared to the transient chaotic sea generated by the homoclinic tangle of the periodic orbit $\gamma$. Then, the volume that escapes from the transient chaotic sea is small in each iteration of the Poincare map, and the unbounded trajectories in the transient chaotic sea intersect the Poincare section many times before escaping to the asymptotic region. 

\begin{figure}[h]
\begin{center}
\includegraphics[scale=0.4]{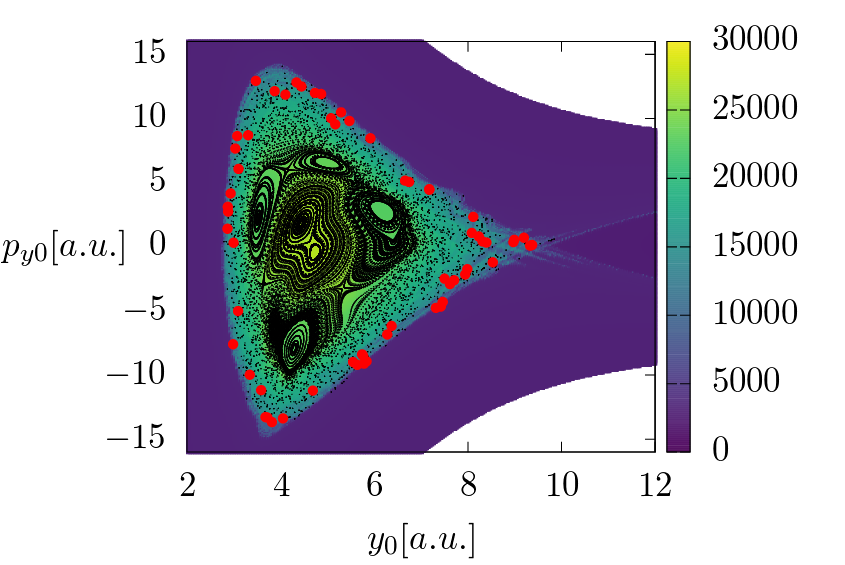}
\caption{Phase space structure of the nonintegrable system for $E = 100 $ K. The red points are the intersections of the trajectory in the figure \ref{fig:trajectories_energies_c} with the Poincare surface. The corresponding Lagrangian descriptor $M_{S_{0}}$ plot for the nonintegrable system is at the background.\label{fig:poincare_nonintegrable}}
\end{center}
\end{figure}

\begin{figure}[h!]
\begin{center}
\includegraphics[scale=0.4]{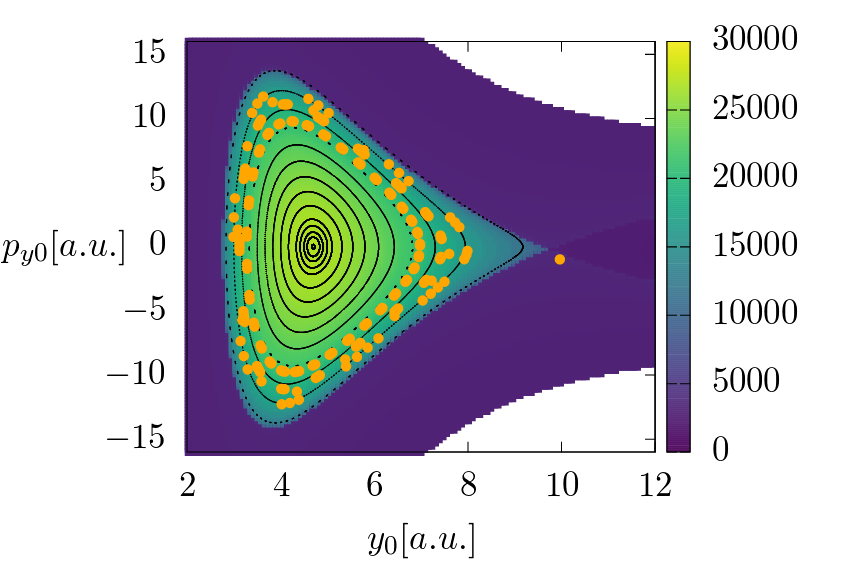}
\caption{The orange points are the iterations of the Poincare map corresponding to the kicked trajectory with $E = 100$ K in the figure \ref{fig:trajectories_energies_c}. This trajectory crosses different invariant closed invariant curves corresponding to the integrable system, the invariant curves are in black. The corresponding Lagrangian descriptor plot for the integrable system is at the background.\label{fig:poincare1_kicked}}
\end{center}
\end{figure}

As we mentioned before, the trajectory evolution rule in the kicked system is the combination of evolution under the influence of potential energy of the integrable system $V_0(\mathbf{r})$ and time-periodic random changes in the direction of the momentum when the trajectory is in the region close to the minimum of $V_0(\mathbf{r})$, $r<5$. To visualise the dynamics generated by the kicks let us consider the Poincare map associated with the orange trajectory in figure \ref{fig:trajectories_energies_e} and the phase space structures in the integrable system generated by $V_0(\mathbf{r})$ for the same value of $E$ as a background. Figure \ref{fig:poincare1_kicked} shows the Poincare map of these orange kicked trajectory, some invariant closed curves of the integrable system, and the Lagrangian descriptor of the integrable system. The random change in the momentum direction is in the interval $[-\pi/12,\pi/12]$. The kicked trajectory intersects with different invariant closed curves in the Poincare section due to the changes in the direction of the momentum and eventually escape from the integrable island.

\begin{figure}[h!]
\begin{center}
\includegraphics[scale=0.4]{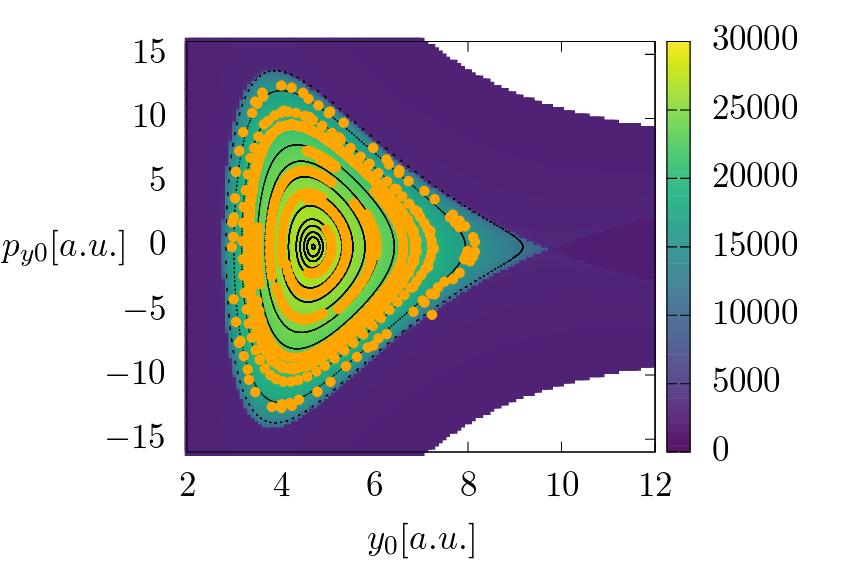}
\caption{The orange points are the iterations of the Poincare map for another the kicked trajectory with $E = 100$ K. This trajectory spends more time in the region defined by the stable island corresponding to the integrable system that the trajectory in the figure \ref{fig:poincare1_kicked}. The corresponding Lagrangian descriptor plot for the integrable system is at the background. \label{fig:poincare2_kicked}}
\end{center}
\end{figure}

The figure \ref{fig:poincare2_kicked} shows the intersections with the same Poincare section of another kicked trajectory with the same initial conditions but a different set of random changes in the direction of the momentum. This trajectory jumps between more invariant closed curves that the trajectory in figure \ref{fig:poincare1_kicked} and spends more time in the integrable island. Nevertheless, both trajectories eventually escape to the asymptotic region.  

\newpage

\section{Escape times from the cauldron}
\label{sec:ET}

In systems with unbounded phase space, a relevant quantity to study is the number of particles that remain in one particular region of the phase space as a function of time, different examples with mixed face space have been studied in \cite{Contopoulosbook,Venegeroles2009}. In the present work, a natural region to consider is the region contained inside the radius of the most external hyperbolic periodic orbit in the phase space. 

For the hyperbolic periodic orbits and its generalisation in more dimensions, the NHIMs, exist a natural surface to study the transport through bottlenecks in phase space. This surface is called the dividing surface associated with the periodic orbit and plays an important role in the transition state theory in phase space proposed by Wigner for systems with two dimensions in \cite{Wigner1938} and extended for systems with more dimensions in \cite{Waalkens2008}. The algorithm to construct a dividing surface of periodic orbits is basically the same as for a NHIM. The procedure consist of three simple steps:

\begin{itemize}

\item Project the periodic orbit (NHIM) in the configuration space.

\item For each point $\mathbf{r}$ in the projection, construct the circumference (sphere) in momentum plane (space) using the equation 

\begin{equation}
\sum_i \frac{p^2_i}{2m_i} = E - V(\mathbf{r})
\end{equation}

\item Take the union of all these circumferences (spheres) in the phase space to construct the dividing surface.

\end{itemize}

The dividing surface associated with a periodic orbit (NHIM) has three important properties in the to study the transport in the phase space and the chemical reaction dynamics:

\begin{itemize}

\item The periodic orbit (NHIM) and its corresponding orbit (NHIM) with opposite momentum are contained in their dividing surface.

\item These two periodic orbits (NHIMs) are the boundaries in the dividing surface between the regions where the trajectories enter into the phase space region contained by the dividing surface and trajectories that left the same region.

\item The flux through the dividing surface is minimal. That is, if the dividing surface of the periodic orbit (NHIM) is deformed, the flux through it increases.

\end{itemize}

For periodic orbits associated with saddle points in the potential energy, the corresponding dividing surfaces are spheres. In the present model, the projection of the hyperbolic periodic orbits $\gamma_0$ and $\gamma$ in the configuration space encircle the potential well then their corresponding dividing surfaces are torus in the phase space. Another recent example of a system with torus genus 1 and 2 as dividing surfaces is in \cite{Montoya2020}. The intersection of the dividing surface with the plane $p_{y0}$--$y_0$ is two vertical segment lines. One segment intersects the periodic orbit $\gamma$, and the other one intersects the periodic orbit with opposite momentum, see figure \ref{fig:poincare_nonintegrable}. All the trajectories that start in the potential well and escape to infinity need to cross the dividing surface. 

\begin{figure}[h!]
\begin{center}
\includegraphics[scale=0.4]{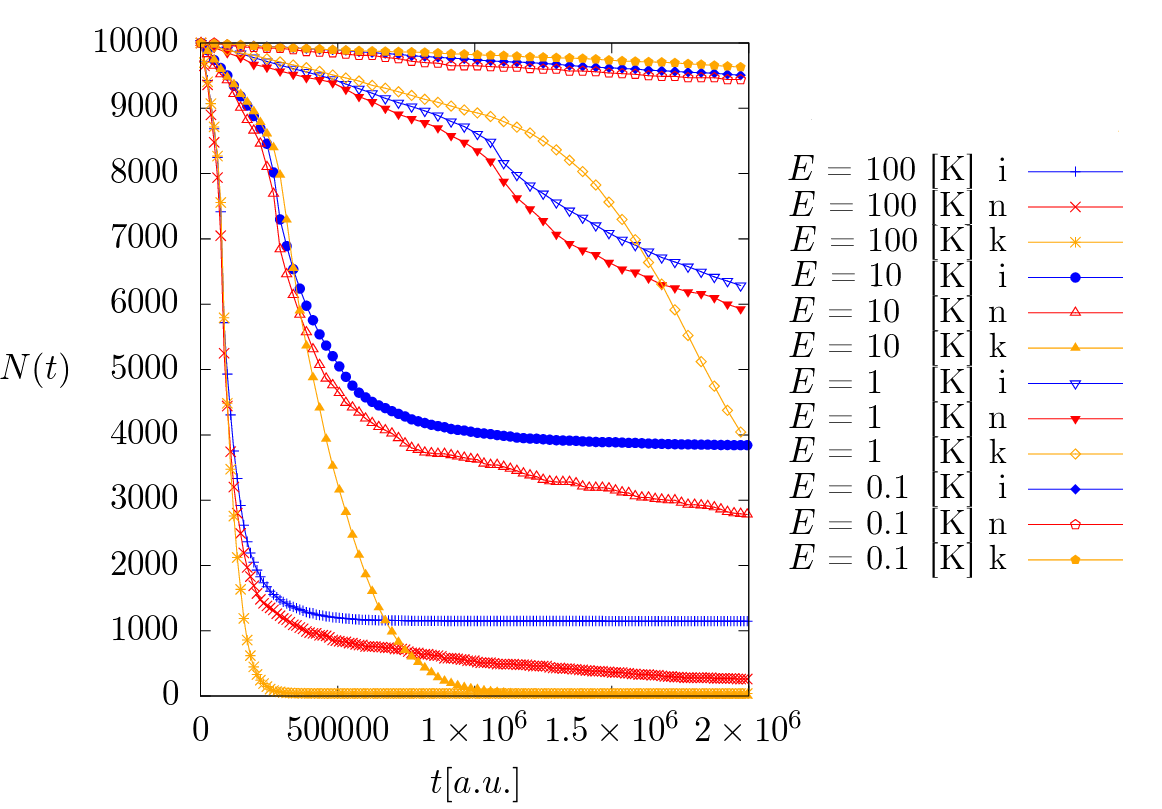}
\caption{Number of particles in the region $R$ as a function of time $N(t)$ for different values of the energy $E$. The time $t$ is in a.u. The blue, orange, and red lines correspond to the integrable, nonintegrable and the kicked systems respectively. \label{fig:N_vs_t}}
\end{center}
\end{figure}

Let us denote by $R$ the region in the constant energy level set delimited by the dividing surface. The intersection of the region $R$ with the plane $p_{y0}$--$y_0$ is the region between the two vertical line segments corresponding to the intersection of the dividing surface with the same plane. The procedure to calculate the number of particles in this region as a function of time $N(t)$ is the following. A random homogeneous distribution of initial conditions with energy $E$ is taken in the region $R$. Their corresponding trajectories are calculated until some maximum time and the number of trajectories that remain in the region $R$ until the time $t$ is recorded. The numerical results for the integrable, perturbed nonintegrable, and kicked systems are in figure \ref{fig:N_vs_t}.

The results show that the behaviour is very similar for the three cases at the beginning. This similarity is related to the blue region shown on the Lagrangian descriptor plots in the figures \ref{fig:LD_integrable_energies} and \ref{fig:LD_nonintegrable_energies}. For the three systems, the dynamics of the trajectories in the blue regions is very similar, except for the trajectories in the small lobes in the perturbed nonintegrable system. However, for larger times, the differences in the curves for $N(t)$ are clear. All the trajectories in the kicked system escape from $R$ and go to infinity, then $N(t)$ goes to zero in a finite time in this case. For the integrable and nonintegrable systems $N(t)$ converge to a constant proportional to the volume of the islands contained in the region $R$. 

\newpage

\section{Conclusions and remarks}

In Hamiltonian systems such that the kinetic energy is a quadratic function of the generalised velocities and potential energy independent of the velocities, it is possible to construct a Lagrangian descriptor $M_{S_0}$ based on the classical action $S_0$. The Lagrangian descriptors are useful tools to reveal the objects in phase space that determine the dynamics, like KAM islands and tangles between stable and unstable manifolds of hyperbolic periodic orbits.

The analysis of the phase space of the variants of this simple 2 dimensional chemical model allows us to understand the differences and similarities of their dynamics and the nature of the escape from the deep potential cauldron that represents the final part of the chemical reaction. The Lagrangian descriptor plots in Fig. \ref{fig:LD_nonintegrable_energies} show how the narrow escape region from the cauldron grows when the energy is increased and the changes of the homoclinic tangle between the stable and unstable manifolds of the most external periodic orbit. For short times the Van der Waals force dominates the motion of the atoms independently of the nature of the short-range interactions between them in the potential cauldron. However, the form of the short interaction is fundamental to explain the long time behaviour of the formation products $N(0)-N(t)$ on the models.
 
In the symmetric case, the most external periodic orbit $\gamma_0$ is associated with the maximum value of the effective potential $V_{0eff}(\mathbf{r})$. The hyperbolic periodic orbit $\gamma_0$ is deformed into the hyperbolic periodic orbit $\gamma$ when the potential energy loses the symmetry, and the system becomes nonintegrable. The periodic orbit $\gamma$ encircles the potential well. Then, the dividing surface for these two degrees of freedom systems is a torus. This is a general property of this type Hamiltonian systems close to a system with rotational symmetry.

In the perturbed nonintegrable version of the model, the additional Gaussian bump in the potential energy breaks the rotational symmetry of the system. The dynamics of the particles with initial conditions in the transient chaotic sea around the KAM islands is determined by the homoclinic tangle between the stable and unstable manifolds $W^{s/u}(\gamma)$. In this example, the exit lobes in the nonintegrable perturbed case are tiny compared with the transient chaotic sea around the KAM islands. Then, the escape time for particles with initial conditions on the transient chaotic sea is larger than for the other particles with initial conditions outside the transient chaotic sea. This scenario is common in open Hamiltonian systems with 2 degrees of freedom and rotational symmetry when a small perturbation breaks the symmetry, and the energy is a little above the threshold energy. The scenario is a consequence of the existence and persistence of the homoclinic tangles of the hyperbolic periodic orbit $\gamma$ and KAM stable islands under perturbations. 

The trajectories in the stable islands remain in the KAM islands all the time for the integrable and perturbed nonintegrable systems. In the case of the kicked system, the momentum kicks make the trajectories jump from one invariant curve to another one. Then, the particles escape from the region defined by the invariant tori of the integrable case to infinity after some finite time. Therefore, the dynamics of the two variants of the model have very different behaviour for long times.

All the previous considerations on the 2 degree of freedom model are the basis for considering the 3 degree of freedom model. In a three degree integrable version, the system has an external NHIM analogous to the most external periodic orbit $\gamma_0$. In this case, this NHIM is the union of all the periodic orbits $\gamma_0$ for all the possible values of vector angular momentum, see \cite{Gonzalez2020,Gonzalez2015,Gonzalez2014} . Their stable and unstable manifolds determine the boundary of the trapped region. The dividing surface is the union of all the dividing surfaces of the periodic orbits $\gamma_0$. When the system is perturbed, and the symmetry breaks, the system becomes chaotic, and the reduction to a two degree of freedom system is not possible. However, the NHIM is robust under perturbation, and their invariant manifolds play an important role in the transport. For short intervals of time, the trajectories in the region close to the NHIM are not significantly affected because of the local nature of the perturbations. The vector field that determines the dynamics is very close to the unperturbed system vector field. The van der Waals force dominates the motion of the particles in the outer region. Further studies are needed to know more properties of 3 degree of freedom model.

\section{Acknowledgments}
\label{sec:Acknowledgments}

We acknowledge the support of EPSRC Grant no. EP/P021123/1. S W acknowledges the support of the Office of Naval Research (Grant No.~N00014-01-1-0769).

\bibliography{biblio}

\end{document}